\documentclass[acmsmall]{acmart}

\AtBeginDocument{%
  }

\setcopyright{acmlicensed}
\copyrightyear{2018}
\acmYear{2018}
\acmDOI{XXXXXXX.XXXXXXX}

\acmJournal{JACM}
\acmVolume{37}
\acmNumber{4}
\acmArticle{111}
\acmMonth{8}

\usepackage{algorithm}
\usepackage{graphicx}
\usepackage{amsmath}
\usepackage{multirow}
\usepackage{booktabs}
\usepackage{array}
\usepackage{subfigure}
\usepackage{amssymb}
\usepackage{makecell}
\usepackage{verbatim}
\usepackage{bm}     
\usepackage{enumitem}       
\usepackage{xcolor}
\usepackage{algpseudocode}
\usepackage{longtable}

\begin{document}

\title{ID-centric pre-training for  recommendation }

\author{Yiqing Wu}

\email{wuyiqing20s@ict.ac.cn}
\affiliation{%
  \institution{Institute of Computing Technology,
Chinese Academy of Sciences and
University of Chinese Academy of Sciences}
  \city{Beijing}
  \state{Beijing}
  \country{China}
}
\author{Ruobing Xie}

\email{ruobingxie@tencent.com}
\affiliation{%
  \institution{Tencent}
  \city{Beijing}
  \state{Beijing}
  \country{China}
}
\author{Zhao Zhang}

\email{zhangzhao2021@ict.ac.cn}
\affiliation{%
  \institution{Institute of Computing Technology,
Chinese Academy of Sciences}
  \city{Beijing}
  \state{Beijing}
  \country{China}
}
\author{Xu Zhang}
\email{zhangzhao2021@ict.ac.cn}
\affiliation{%
  \institution{Institute of Computing Technology,
Chinese Academy of Sciences}
  \city{Beijing}
  \state{Beijing}
  \country{China}
}
\author{Fuzhen Zhuang}
\email{zhangzhao2021@ict.ac.cn}
\affiliation{%
  \institution{Institute of Artificial Intelligence,
Beihang University}
  \city{Beijing}
  \state{Beijing}
  \country{China}
}
\author{Leyu Lin}
\email{goshawklin@tencent.com}
\affiliation{%
  \institution{Tencent }
  \city{Beijing}
  \state{Beijing}
  \country{China}
}
\author{zhanhui Kang}
\email{kegokang@tencent.com}
\affiliation{%
  \institution{Tencent }
  \city{Beijing}
  \state{Beijing}
  \country{China}
}
\author{Yongjun Xu}
\email{xyj@ict.ac.cn}
\affiliation{%
  \institution{Institute of Computing Technology,
Chinese Academy of Sciences }
  \city{Beijing}
  \state{Beijing}
  \country{China}
}
\

\renewcommand{\shortauthors}{Yiqing Wu et al.}

\begin{abstract}
  Classical sequential recommendation models generally adopt ID embeddings to store knowledge learned from user historical behaviors and represent items. However, these unique IDs are challenging to be transferred to new domains. With the thriving of pre-trained language model (PLM), some pioneer works adopt PLM for pre-trained recommendation, where modality information (e.g., text) is considered universal across domains via PLM. Unfortunately, the behavioral information in ID embeddings is still verified to be dominating in PLM-based recommendation models compared to modality information and thus limits these models' performance.
In this work, we propose a novel ID-centric recommendation pre-training paradigm (IDP), which directly transfers informative ID embeddings learned in pre-training domains to item representations in new domains. Specifically, in pre-training stage, besides the ID-based sequential model for recommendation, we also build a Cross-domain ID-matcher (CDIM) learned by both behavioral and modality information. In the tuning stage, modality information of new domain items is regarded as a cross-domain bridge built by CDIM. We first leverage the textual information of downstream domain items to retrieve behaviorally and semantically similar items from pre-training domains using CDIM. Next, these retrieved pre-trained ID embeddings, rather than certain textual embeddings, are directly adopted to generate downstream new items' embeddings. Through extensive experiments on real-world datasets, both in cold and warm settings, we demonstrate that our proposed model significantly outperforms all baselines. Codes will be released upon acceptance.
\end{abstract}


\received{20 February 2007}
\received[revised]{12 March 2009}
\received[accepted]{5 June 2009}

\maketitle

\section{Introduction}
Sequential recommendation (SR) takes user‘s historical behavior sequence chronologically as input and outputs the next predicted items according to user preference \cite{kang2018self,sun2019BERT4Rec} and is applied to websites' recommender systems. Classical SR models mostly rely on ID embeddings to represent items (i.e., projecting the unique ID of an item into a dense embedding). 
\begin{figure}[!hbtp]
\centering
\includegraphics[width=0.70\textwidth]{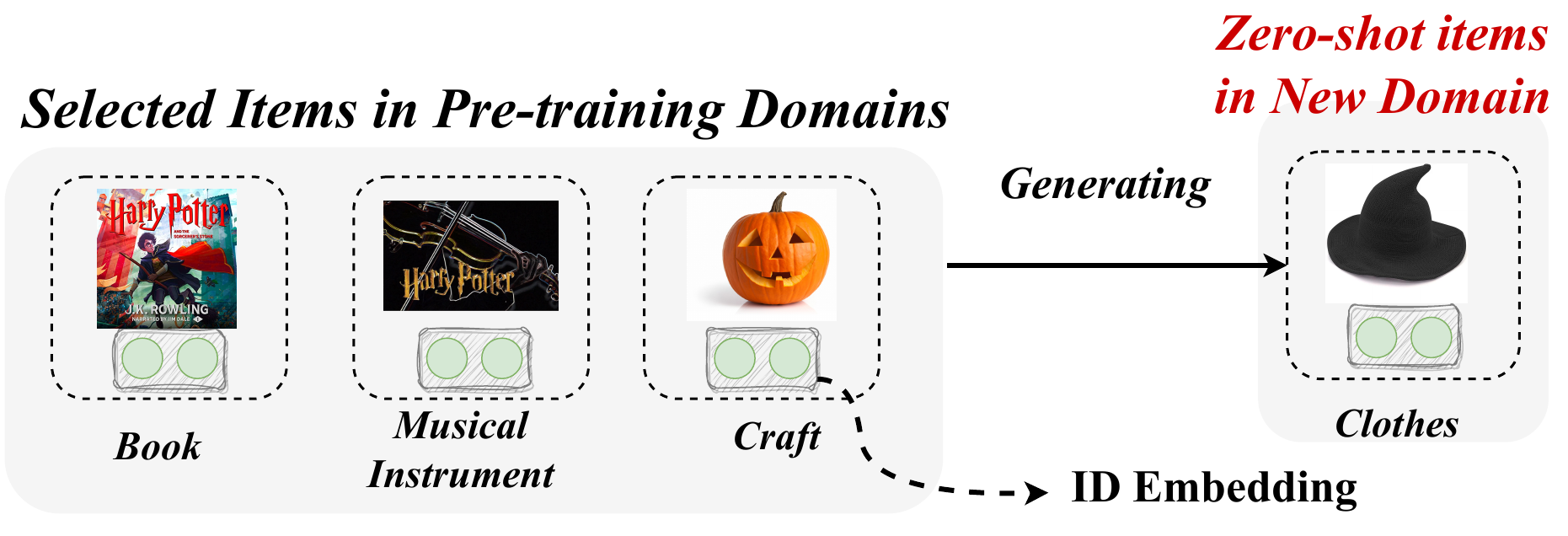}
\caption{A example of ID-centric pre-training recommendation. The central idea is to select related pre-training ID embeddings to generate item embeddings in new domains.}
\label{fig:example}
\end{figure}
Undoubtedly, the behavioral information hidden in user-item interactions forms the foundation of recommendation. ID embeddings + ID-based sequential model (e.g., \cite{kang2018self}) are perfectly suitable for storing such information in SR and are dominating in practical SR systems.
However, such an ID-centric SR model cannot well deal with the cross-domain recommendation, since the well-learned unique ID embeddings are challenging to directly transfer to downstream new domains. 
Some efforts have explored cross-domain recommendation \cite{hu2018conet,tsai2018diversity, Zhu_Chen_Wang_Liu_Zheng,zhu2022personalized},
while most of them rely on certain overlapping users or items with customized networks, which are limited and not that universal.
In fact, the untransferability of ID embeddings is becoming a stumbling block in building a universal SR model for multiple domains. 


To overcome the untransferability of ID embeddings, some works attempt to bypass IDs with modality information (e.g., texts).
Recently, pre-trained models have achieved substantial advancements  CV and NLP fields \cite{devlin2019bert,mann2020language, Liu_Lin_Cao_Hu_Wei_Zhang_Lin_Guo_2021}. 
Inspired by this, a series of pioneering initiatives have endeavored to utilize the abundant knowledge of pre-trained language models (PLM) to address cross-domain adaptation and cold-start problems in recommendation. Some works infer user preferences from textual modality information \cite{hou2022towards,li2023gpt4rec}.
For example, UniSRec \cite{hou2022towards} relies on item textual information with PLM and parameter whitening to represent items in multiple domains. Some studies further adopt PLM for sequential behavior modeling \cite{cui2022m6,zhang2023recommendation}.
However, some works demonstrate that there still exists a huge gap between PLM (textual information) and recommendation (behavioral information) \cite{gao2023chat,liu2023chatgpt,hou2023large}. The help of common knowledge in PLMs is marginal when we have sufficient user-item interactions.

Despite the remarkable progress achieved by PLM-centric recommendation models, we conclude three key factors that constrain the practical usage of existing PLM-based recommendation models for multi-domain recommendation:
\textbf{(1)} Behavioral information in ID embeddings has been verified to be dominating for SR compared to modality information, especially for relatively dense datasets. The static common knowledge in PLM cannot directly infer dynamic personalized preferences in SR.
Sometimes even a well-tuned single-domain SR model (e.g., SASRec \cite{kang2018self}) can achieve better results \cite{gao2023chat}. Modality representations cannot fully replace ID embeddings, and directly representing items with texts may hinder sufficient model training.
\textbf{(2)} Item modality information may contain noise and conflicts, which cannot accurately and comprehensively represent items (i.e., textual correlation does not always imply behavioral correlation). Even items with the same modality information may represent different behavior patterns.
\textbf{(3)} Currently, industrial recommendation models are mostly based on IDs. Considering the efficiency and costs, PLM-based models are too heavy and cost-intensive in practical systems.
Despite the great temptation of using LLM to ``conquer'' recommendation, we still argue that: \textbf{ID-based sequential modeling} is still the best practical and most efficient choice for recommendation currently, while modality information is more suitable as supplementary features for zero-shot scenarios.


Based on the above observations, we believe that \textit{establishing an ID-centric Pre-trained Recommendation Model is crucial.} Unfortunately, there have been no prior attempts in this regard, and we aim to take the first step. To achieve this, the core problem is \emph{how to effectively transfer the informative knowledge encoded in pre-trained ID-based sequential model to new domains}.
Naturally, item modality information (e.g., texts) widely exists and is relatively universal across domains, which could be regarded as a bridge to connect multi-domain ID embeddings in pre-training and downstream tasks as shown in Fig.\ref{fig:example}. 
To achieve this, we design a novel \textbf{ID-centric pre-training framework (IDP)} for multi-domain recommendation. Specifically, we first optimize an ID-based sequential model (e.g., SASRec) on all interactions in multiple pre-training datasets. Besides, since textual similarity does not always equal behavioral similarity, we also build a \textbf{cross-domain ID matcher (CDIM)} to connect multi-domain item IDs, which considers both behavioral and textual modality correlations. In tuning, for each new item, we first retrieve its top (behaviorally and textually) similar items in pre-training datasets via item's textual information via CDIM, and then aggregate their ID embeddings as the new item's initialization.
Overall, the contributions of this work are summarized as follows:
\begin{itemize}
    \item To the best of our knowledge, we are the first to build an ID-based pre-trained model for multi-domain recommendation. It creatively represents new-domain items via pre-trained ID embeddings, directly using IDs for main recommendation and modality for multi-domain ID connections.
    \item We implement IDP via CDIM, which utilizes behavior-tuned textual correlations as bridges to aggregate well-learned ID embeddings for new items. CDIM adopts the universal textual information to reuse pre-trained knowledge while alleviating too many behavior-text conflicts.
    \item Extensive experiments on $4$ datasets have verified the effectiveness and universality of our ID-based pre-training. IDP is effective, efficient, easy-to-deploy, and universal, which is welcomed to be quickly evaluated in the industry.
\end{itemize}
\section{Related Works}
\label{sec.related_work}

\noindent
\subsection{Sequential Recommendation}
Sequential recommendation models mainly leverage users' chronological behavior sequences to learn users' preferences.
Early works model sequence patterns by Markov chains (MCs). Shani et al. \cite{shani2005mdp} formalized sequential recommendation as a sequential decision problem and adopted Markov Decision Processes (MDPs) to solve it. Later \cite{rendle2010factorizing} combine the power of MCs and MFs and achieve a promising result. High-order sequential relations also be considered in the following work\cite{he2016fusing,he2017translation}. Recently, with the great progress of deep learning,  ID-based sequential models have achieved great success, which project item ID to unique dense embedding. 
GRu4Rec introduces the GRU network to session-based recommendation for ranking tasks. On the top of GRU4Rec, memory networks\cite{chen2018sequential,huang2018improving}, hierarchical structures\cite{quadrana2017personalizing}, attention mechanisms \cite{liu2018stamp}, reinforcement learning \cite{xin2020self} .etc  are introduced. Expected recurrent neural network, other models also employed for sequential recommendation. Caser \cite{tang2018personalized} regards the user behavior sequence as a one-dimensional image and adopts the Convolutional Neural Network (CNN) to capture sequence patterns.

Inspired by the success of Transformer and BERT \cite{vaswani2017attention,devlin2019bert}, attention and transformer-based models are introduced to the recommendation. DIN \cite{zhou2018deep}  applies attention mechanisms between the target item and user behavior sequence to capture the user interest.
SASRec \cite{kang2018self} utilizes the single-directional Transformer block to model the sequential.  Following BERT\cite{devlin2019bert}, BERT4Rec \cite{sun2019BERT4Rec} adopt a bidirectional Transformer to incorporate user behavior information from both directions. FDSA\cite{zhang2019feature} considers the features of items and designs
 a feature-level self-attention
block to leverage the attribute information about items in the user
behavior sequence.


\subsection{Pre-training in Recommendation.} 

Recently, pre-training models have achieved great success in NLP \cite{devlin2019bert,liu2019roberta} and CV \cite{he2021masked,radford2021learning}. They aim to transfer prior knowledge from general large-scale datasets to downstream domains. While in recommendation, since IDs are unaligned, the well-learned unique ID embeddings are challenging to transfer directly to downstream new domains.  
To utilize the power of pre-trained model, some single-domain pre-trained models \cite{yao2021self,wu2021self} have been proposed to improve recommendation accuracy of the inner domain. However, these models lack transferability and are unable to apply to cross-domain recommendation. Some works transfer well-learned knowledge by overlapped users/items \cite{hu2018conet,tsai2018diversity, Zhu_Chen_Wang_Liu_Zheng,zhu2022personalized,zang2022survey,Yuan_Zhang_Karatzoglou_He_Jose_Kong_Li_2020}. While overlapped users/items generally are limited in the real world, thus they are lack of universality. To build a universal model, some works attempt to design a text-based/modality-based pre-trained model \cite{hou2022towards,ding2021zero,cui2022m6}. Those works introduce pre-trained language models (PLM) into recommendations and attempt to replace ID-embedding with modality information.  Specifically, UniSRec \cite{hou2022towards} utilizes PLM to learn universal item representations across different domains via items' textual information. However, recent studies show that ID-embedding is still irreplaceable \cite{gao2023chat,liu2023chatgpt,hou2023large}. Different from previous works, we build an ID-centric pre-trained model, which directly transfers pre-trained ID-embedding to downstream domains.

\section{Methodology}
\begin{table}
  \caption{Mathematical Notations}
  \label{tab:symbol}
  \begin{tabular}{cl}
  \toprule
    Symbols & Description\\
    \midrule
    $S$, $T$ & The pre-training multi-domain dataset and downstream domain \\
    $U$, $V$ &  The User set of  and the Item set \\
    $u$, $v$ & The user and item ID   \\
    $s_u$ &  The behavior sequence of user $u$    \\
    $E, \mathbf{e_v}$ & The Item ID embedding Matrix and the ID embedding of item $v$\\
    $P,\mathbf{p_i}$ & The Position Embedding Matrix and the position embedding of index $i$ \\
    &in behavior sequence.\\
    $\mathbf{e_u}$ & The encoded  user embedding\\
    $\mathbf{h}_u^0$ & The input representation of sequential encoder\\
    $Trm$ & The transformer block \\
    $SimCSE$ & The encoder of SimCSE\\
    $c_{v}$ & the textual information of item $v$\\
    $t_v$  & the encoded textual representation of encoded item \\
    $m, TOP(\cdot,\cdot)$ & The number of selected similar items of Pre-train\\
     &domain and the function that finds top-m items \\
    
      $V_{v^T}$ & The set of selected top-m similar items of the Pretraning domain\\

      $f_{seq}$ & The sequential encoder of pre-training\\
      $f_{seq}^{new}$ & The sequential encoder that re-trained on target downstream domain\\
      $\Theta, \Theta^*$ & The parameters of the pre-trained sequential encoder and  the parameters  \\&
      of fine-tuned sequential encoder  \\
      $\vartheta^*$ & The parameters of re-trained new sequential encoder by target downstream \\
      &  domain  \\

  \bottomrule
\end{tabular}
\end{table}

\subsection{Problem Statement}

 In this paper, we denote the user and item as $u \in U$ and $v \in V$, where $U$ and $V$ are the user set and item set. Each item is  associated with a unique ID $v_i$ and has a piece of text $c_{v_i}=\{w_1,w_2, ..., {w_{|c_{v_i}|}}\}$ that describes it, where $w_j$ is the natural word in language. 
Besides, each user has a corresponding chronological behavior sequence  $s_u=\{v_1^u,v_2^u,...,v_{|s_u|}^u\}$, where $|s_u|$ is the length of user $u$'s  behavior sequence.  Each item $v_i$ is associated with a trainable ID embedding denoted as  $e_{v_i}$. We use $S$ and $T$ to denote the pre-training multi-domain dataset and downstream domain dataset, respectively. The items and users in pre-training domain and target domain are denoted with $v_i^S$,$u_i^S$, $v_i^T$, and $u_i^T$, respectively.

Sequential Recommendation aims to predict appropriate items that the user would like to interact with next time $|s_u|+1$ based on the user behavior sequence $s_u$. It can be formulated as follows:
\begin{equation}
    v_i^*= \arg \max_{v_i\in V} P(v_{|s_u|+1}=v_i|s_u).
\end{equation}

\subsection{Overview}

\begin{figure*}[!hbtp]
\centering
\includegraphics[width=1.0\textwidth]{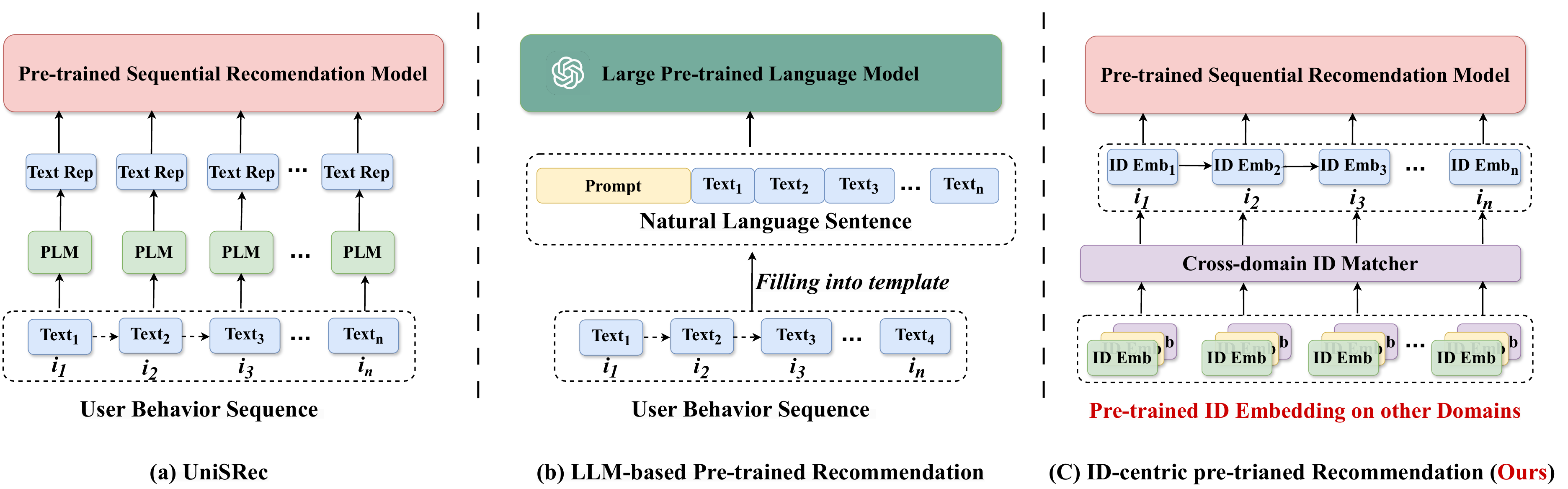}
\caption{Illustrations of our IDP and other pre-trained models. Different from conventional pre-trained models that use texts (representations) as items for behavior modeling, our IDP directly uses pre-trained ID embeddings to generate new items.}
\label{fig:VS}
\end{figure*}

In this paper, we aim to propose an ID-centric pre-trained recommendation model, called IDP. We compare our IDP with other methods and show it in Fig.\ref{fig:VS}. The key difference between our approach and previous works lies in we do not attempt to replace IDs with textual modality information. Instead, our IDP directly transfers the behavioral information inherent in the upstream ID embeddings to the downstream. We argue that ID-based recommendation still dominates recommendation, and our IDP provides a new path to build a pre-trained recommendation model.
 
\begin{figure*}[!hbtp]
\centering
\includegraphics[width=0.75\textwidth]{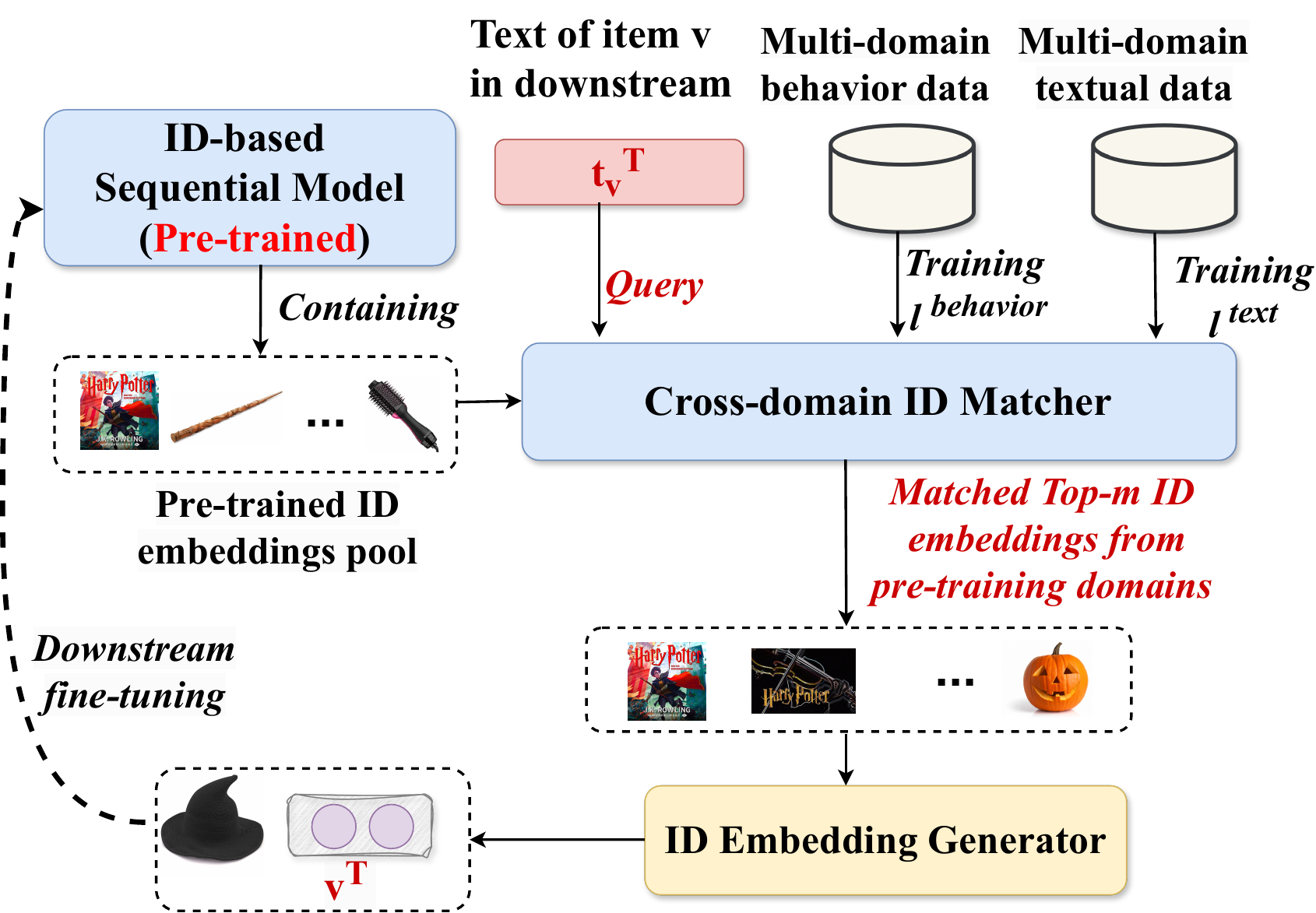}
\caption{Overall architecture of our IDP.}
\label{fig:Matcher}

\end{figure*}

The framework of IDP is illustrated in Fig.\ref{fig:Matcher}.  Our IDP mainly includes three parts: (1)  ID-based sequential model pre-training (2) Cross-domain ID Matcher  (3) downstream fine-tuning. Specifically, we first pre-train an ID-based sequential model on a multi-domain pre-training dataset. Subsequently, in order to model the behavior-aware item modality correlations in recommendation, we train a Cross-domain ID Matcher (CDIM) with collaborative signals. With the help of CDIM, we can leverage textual information as a bridge to connect downstream items with pre-trained item embeddings. After that, we utilize the selected pre-trained item embeddings obtained by CDIM to generate ID embeddings for the downstream domain. Lastly, we fine-tune both the pre-trained model and ID-generated ID embeddings for the new domain task. Through this framework, our IDP could handle multiple new domains while taking advantage of powerful and efficient ID-based sequential modeling. We summary the



\subsection{Pre-trained ID-based Sequential Model}

In this section, we introduce the pre-training stage of our IDP. Sequential recommendation has been widely studied in previous works \cite{kang2018self,sun2019BERT4Rec}. Following \cite{zhou2020s3,xie2022contrastive,hou2022towards}, we adopt the classical SASRec \cite{kang2018self} as our base sequential model.  Note that our IDP is flexible and
does not depend on specific sequential models, it can be easily applied to other sequential models. We also evaluate it in the experiment section.
Specifically, for a user sequence $s_u=\{v_1^u,v_2^u,...,v_{|s_u|^u}\}$, We input it to the sequential model and it can be formulated as:
\begin{equation}
\begin{aligned}
\mathbf{e}_u=f_{seq}(s_u|\Theta, \mathbf{E}, \mathbf{P}), where \, f_{seq} \, is  \, SASRec,
\end{aligned}
\end{equation}
For $l=0$, $h_{u, i}^0$ is the ID embedding of $v_i^u$.

Following classical recommendation models \cite{kang2018self,xie2022contrastive}, we use BPR loss \cite{rendle2009bpr} to optimize our pre-training model, formulated as:
\begin{equation}
\begin{split}
L_o = -\sum_{(u,v_i) \in S^{+}} \sum_{(u,v_j) \in S^{-}} \log \sigma (\mathbf{e}_u^\top\mathbf{e}_{v_i} - \mathbf{e}_u^\top\mathbf{e}_{v_j}), \quad u\in U^S, v_i,v_j
\in V^S.
\end{split}
\label{eq.bpr_loss}
\end{equation}
We pre-train IDP on multi-domain datasets by simply merging all domains' interactions without any user/item alignment. 
\subsection{Cross-domain ID Matcher}
\label{secCDIM}

For building an ID-based pre-training model, the biggest challenge lies in the misalignment between the IDs in the pre-training multi-domain and those in the downstream domain, which makes it impossible to directly utilize the upstream ID embeddings.
However, we notice that, in most systems, items have modality information (e.g., title, content). Generally, the modality information is universal across domains. Hence, we hope to utilize textual information as a bridge to build a Cross-domain ID Matcher (CDIM). CDIM can match items in downstream domains to appropriate items in pre-training domains.  In this way, we can utilize the pre-trained ID embedding for downstream tasks. 

A straightforward solution is to utilize general textual modality similarity for identifying similar items. However, merely considering modality similarity falls short of achieving a proficient CDIM. There exist gaps between the general semantic space and that in recommendation. Actually, in recommendation, behavioral information is crucial and should also be taken into account when measuring similarity.
For example, paper diapers and beer may not exhibit semantic similarity in the general semantic space, but they might be considered similar in a recommendation context since new fathers often purchase them together.
Overall, we believe a proficient CDIM  should take into account two fundamental factors: \textbf{\textit{(1) textual similarity.  (2) behavior similarity.  }}
\subsubsection{Textual similarity }

Actually, measuring the similarity of two items for textual modality is similar to matching two sentences.  Previous works directly adopt BERT \cite{hou2022towards} as the textual information encoder, while BERT is not designed for the sentence-matching task.
Considering this, we utilize SimCSE \cite{gao-etal-2021-simcse} to encode the textual information. SimCSE employs contrastive learning to enable the BERT model \cite{devlin2019bert} to generate high-quality sentence embeddings, resulting in excellent performance on the sentence-matching task. Specifically, given an item $v_i$ and corresponding textual information $c_{v_i}=\{w_1,w_2, ..., {w_{|c_{v_i}|}}\}$, we first concatenate a special token $[CLS]$ with $c_{v_i}$, then we feed the concatenated text into SimCSE as:
\begin{gather}
    t_{v_i} = SimCSE(\{[CLS],w_1,w_2, ..., {w_{|c_{v_i}|}}\}).
\label{eq.SimCSE_encoder}
\end{gather}
$t_{v_i}$ is the representation of $[CLS]$ token from the last layer of $SimCSE$. 

Furthermore, considering that SimCSE is training on the open-accessed corpus, and may exist a gap between the recommendation corpus. We further fine-tune SimCSE on recommendation data with a contrastive learning task. Similar to SimCSE, we simply feed the same input textual information to the encoder twice.  In this way, we get two vectors $t_{v_i}$ and $t_{v_i}^{'}$, which are encoded by two randomly dropped encoders (the encoder has dropout operation), respectively. We treat $t_{v_i}$ and $t_{v_i}^{'}$ as a positive pair.  The contrastive learning task can be formulated as:
\begin{equation}
\begin{split}
   l_{v_i}^{text}=\frac{e^{sim(t_{v_i},t_{v_i}^{'})/\tau}}{e^{sim(t_{v_i},t_{v_i}^{'})/\tau}+\sum_{k}^{N-1} e^{sim(t_{v_i},t_{v_k})/\tau} }.
\end{split}
\label{eq.SimCSE_textual}
\end{equation}
where $t_i$,$t_i^{'}$ is the corresponding  textual vectors encoded by $SimCSE$, $\tau$ is the temperature parameter. 
$N$ is the batch size, we adopt an in-batch negative sample strategy, where the other items' textual vectors $t_k$ are regarded as negative instances.  
\subsubsection{Behavior-involved tuning}
As discussed before,  behavioral information is crucial and should also be taken into account when measuring similarity in recommendation. 
Inspired by the contrastive learning task of $SimCSE$, we utilize collaborative signals to help model learning the behavior similarity in the recommendation scenario. Specifically, we first train a recommendation model on \textbf{multi-domain pre-training datasets}. In this work, we directly use the pre-trained sequential model in the last step. Subsequently, for each item $v_i$ in pre-training datasets, we obtain its most similar items $V_s^{v_i}=(v_{s,1}^{v_i},v_{s,2}^{v_i}...,v_{s,k}^{v_i}) $ in pre-training datasets by embedding similarity of pre-trained model.  Last we conduct contrastive learning to fine-tune SimCSE. Here, We regard $<v_i,v_j>, v_j \in V_s^{v_i}$ as a positive pair and $<v_i,v_k>, v_k \notin V_s^{v_i}$ as a negative pair. For an item $v_i$, the contrastive learning task  can be formulated as:
\begin{equation}
\begin{split}
   l_{v_i}^{behavior}=\sum_{v_j \in V_s^{v_i}}\frac{e^{sim(t_{v_i},t_{v_j})/\tau}}{e^{sim(t_{v_i},t_{v_j})/\tau}+\sum_{k}^{N-1} e^{sim(t_{v_i},t_{v_k})/\tau} } .
\end{split}
\label{eq.SimCSE_behavior}
\end{equation}

We adopt the multi-task learning strategy to fine-tune SimCSE, so as to make the similarity of ID embeddings able to reflect both semantic (used to connect multi-domain items via texts) and behavior (better for new-domain initialization in recommendation) correlations. Specifically, the loss can be represented as:
\begin{equation}
\begin{split}
   L=\sum_{v_i \in V} l_{v_i}^{behavior} + l_{v_i}^{text}.
\end{split}
\label{eq.SimCSE}
\end{equation}

\subsection{Generate ID Embedding in New Domains}

In this section, we introduce how we use the pre-trained item ID embedding to generate the item ID embedding of downstream scenarios.
With the help of CDIM, we can measure the similarity of items by textual information. Intuitively, items that are similar may display analogous interaction patterns in the recommendation, which is stored in ID embedding. 
Hence, for an item in the downstream domain, we leverage the item's textual information as a bridge to connect this item with related pre-trained item ID embeddings to generate its ID embedding. This process can be formulated as follows:
\begin{equation}
    \begin{split}
    V_{v_i^T}=TOP([sim(t_{v_i}^T,t_1^S),sim(t_i^T,t_2^S),...,sim(t_i^T,t_{|V^S|}^S)],m),
    \end{split}
\label{eq.TOPKitems}
\end{equation}
where $TOP(\cdot,\cdot)$ is the function that finds the top-$m$ items by similarity scores given by CDIM, $V_{v_i^T}$ is the top-$m$ similar items. Note that this process can be fast, we have analyzed the complexity of our model in \textbf{Sec. \ref{sec:complexity}.}

After that, we utilize the corresponding ID embeddings of the top-$m$ items to generate an ID embedding for the item $v_i^T$ in the target domain. 
Considering that, for an item $v_i^T$ in the target domain, the significance of its similar items in the pre-trained domains may vary based on their similarity to $v_i^T$. Therefore, for a given item $v_i^T$ and its corresponding top-$m$ most similar items ${V_{v_i^T}=\{{v_1^S, v_2^S, ..., v_m^S}}\}$ in the pre-training domains, we employ normalized similarity as attention weights to generate the ID embedding for $v_i^T$.
\begin{equation}
\begin{split}
    score_{{v_i},{v_j}}&=sim(t_{v_i}^T,t_{v_j}^S) / \sum_{k}^{m}{sim(t_{v_i}^T,t_{v_k}^S)},\\
   e_{v_i}^T&=\sum_k^m{score_{i,k}*e_{v_k}^S},
\end{split}
\label{eq. generate embedding}
\end{equation}
where $e_k^S$ is the ID embedding of $v_k^S$, $e_{i}^T$ is the generated ID embedding for $v_i^T$ in the downstream.

\subsection{Applied to downstream domain}
After generating ID embeddings for the target domain, we can apply the pre-trained model to downstream domains. Different from previous methods that are bonded with a pre-trained encoder, our IDP focuses on transfer the knowledge contained in ID embedding. Thus our IDP is fixable It can be applied in the following ways:
\begin{itemize}
     \item We can directly apply the model in downstream domains. In this setting, we directly deploy the generated ID embedding and the pre-trained sequential model to  downstream domains without any tuning. This process can be formulated as :
     \begin{equation}
\begin{split}
    \mathbf{e}_u^T=f_{seq}(s_u^T|\Theta,\mathbf{E}^T, \mathbf{P}),
\end{split}
\label{eq. generate embedding1}
\end{equation}
    where the $f_seq$ is the pre-trained sequential model, $\Theta$ is the parameters of the pre-trained model, and it contains the parameters of the embedding layer and sequential encoder .
    \item we can fine-tune both pre-trained sequential encoder and generated ID embedding on downstream data. This can be formulated as:
    
\begin{equation}
\centering
\begin{split}
     \mathbf{e}_u^T&=f_{seq}(s_u^T|\Theta^*, \mathbf{E^*}^T, \mathbf{P}^*),\\
     \Theta ^*, \mathbf{E^*}^T, \mathbf{P}^* &=\arg \max_{\Theta,\mathbf{E}^T, \mathbf{P}}P(v_{|s_u|+1}|s_u,\Theta), u\in U^T, v\in V^T,
\end{split}
\label{eq. generate embedding2}
\end{equation}
    where $\Theta ^*$ is the fine-tuned parameters.
    \item We can re-train another encoder (it can be a different architecture from the pre-trained model) and fine-tune generated embedding on downstream data. 
    \begin{equation}
\centering
\begin{split}
     \mathbf{e}_u^T&=f_{seq}^{new}(s_u^T|\vartheta^*, \mathbf{E^*}^T, \mathbf{P}^*),\\
     \vartheta ^*, \mathbf{E^*}^T, \mathbf{P}^* &=\arg \max_{\vartheta, \mathbf{E}^T, \mathbf{P}}P(v_{|s_u|+1}|s_u,\vartheta), u\in U^T, v\in V^T,
\end{split}
\label{eq. generate embedding3}
\end{equation}
 where $\vartheta ^*$ is the parameters of the new sequential model.
 \end{itemize}

\subsection{Fine-tuning Pre-trained Model}
 After generating ID embeddings for the target domain. We optimize pre-trained model with the supervision of the target domain's behavior data. Different from previous methods \cite{hou2022towards} that are bonded with the pre-trained encoder, our IDP is flexible as we transfer knowledge by ID embedding. Our IDP can be applied  in the following ways:
 \begin{itemize}
     \item we can directly deploy the model in downstream domains with the generated new ID embedding and pre-trained sequential encoder.
 \end{itemize}
 (1) we can directly deploy the model in downstream domains with the generated new ID embedding and pre-trained sequential encoder. (2) we can fine-tune both pre-trained sequential encoder and generated ID embedding on downstream data. (3) We can re-train another encoder (it can be a different architecture from the pre-trained model) and fine-tune generated embedding on downstream data. 
 This process can be formulated as follows:
\begin{equation}
\begin{split}
    \Theta ^* =\arg \max_{\Theta}P(v_{|s_u|+1}|s_u,\Theta), u\in U^T, v\in V^T,
\end{split}
\label{eq. fine-tuning}
\end{equation}
where $\Theta$ is the parameters of fine-tuning, which varies depending on the fine-tuning approach.
\begin{algorithm}

    \caption{Training process of IDP}\label{alg:adaptive_prediction}
    \begin{algorithmic}[1] 
    \Require Pre-training multi-domains  user set $\bm{U^S}$ and  item set $\bm{V^S}$;Pre-training dataset $\mathcal{X}^T$ ; Pre-trianing Sequential recommendation encoder $\bm{f_{seq}}$; Target downstream domain user set $\bm{U^T}$ and target downstream domain user set $\bm{V^T}$; Target downstream data $\mathcal{X}^T$
    \\
    \State$\mathbf{\Theta}$, $\mathbf{E}^S$, $\mathbf{P}$,$f_{seq} \gets$ Call function: \textbf{\MakeUppercase{Pre-training sequential model}}
    \State Call function: \textbf{\MakeUppercase{Training CDIM}}
     \State$\mathbf{E}^S\gets$ Call funciton: \textbf{\MakeUppercase{ Generating ID embedding}}
    \State Fine-tuning by \textbf{Eq.\ref{eq. generate embedding1},Eq.\ref{eq. generate embedding2} or Eq.\ref{eq. generate embedding3}}
    \\
    \Function {Pre-training sequential model}
    {}\label{module:reviewimitator}
    \State Initialize the trainable parameters $\mathbf{\Theta}$ of encoder $f_{seq}(\cdot)$, item embedding matrix $\mathbf{E}$ and position matrix $\mathbf{P}$;
     \Repeat
     \For {$\textit{batch data}$ \textbf{in} $\mathcal{X}^S$ }
     \State $\mathbf{e}_u \gets $  encode the user behavior sequence $f_{seq}$(\textit{bath data}) 
      \State $L_0 \gets$ get the loss with Eq.\ref{eq.bpr_loss} 
      \State \textbf{Update} $\mathbf{\Theta}$, $\mathbf{E}$ and $\mathbf{P}$ according $L_0$
     \EndFor
      \Until{the trainable parameters have converged.}
      \State   \textbf{return} {The pre-trained $\mathbf{E}$, $\mathbf{P}$, $\mathbf{\Theta}$ and $f_{seq}$}
    \EndFunction
    \\
    \Function {Training CDIM}{} \label{module:CDIM}
     \Repeat
     \State $L^{text} \gets$ calculate the textual contrastive learning loss with Eq.\ref{eq.SimCSE_textual}
     \State $L^{behavior} \gets$ calculate  behavior-involved contrastive learning loss with Eq.\ref{eq.SimCSE_behavior}
     \State $\L \gets L^{text}+L^{behavior}$
     \State \textbf{Update} SimCSE by $L$ 
     \Until{the trainable parameters have converged. }
    \EndFunction
        \\
    \Function {Generating ID embedding}{} \label{module:GE}
     \State$V_{v_i^T} \gets$ get the most similar items in $V^S$ by Eq.\ref{eq.TOPKitems} 
     \State $E^T\gets$ generate item embedding with Eq.\ref{eq. generate embedding} 
    \EndFunction

    \end{algorithmic}
\end{algorithm}
\subsection{Base Model}
In this paper,  Following \cite{zhou2020s3,xie2022contrastive,hou2022towards}, we adopt the classical SASRec \cite{kang2018self} as our base sequential model. 
SASRec encodes the user behavior sequence by stacking the transformer layers and achieving promising performance. The SASRec consists of three sub-modules: an embedding layer, a multi-head attention module and a position-wise feed-forward Network.
Note that our IDP is flexible and does not depend on specific sequential models, it can be easily applied to other sequential models.
\subsubsection{Embedding Layer}
The embedding layer projects the high-dimensional one-hot item ID to low-dimensional dense embedding by looking up an item embedding table $\mathbf{E} \in \mathbb{R}^{|V| \times d}$. In addition, there is also a position embedding table $\mathbf{P} \in \mathbb{R}^{T \times d}$ that projects the position of items in sequence to dense embedding, where  $T$ is the maximum length of the sequence. Given a user behavior sequence $s_u=\{v_1^u,v_2^u,...,v_{|s_u|}^u\}$, we can obtain the input representation of user behavior sequence:
\begin{equation}
\begin{split}
  \mathbf{h}_u^{0}&=\{\mathbf{h}_{u,1}^{0},\{\mathbf{h}_{u,2}^{0},...,\mathbf{h}_{u,|s_u|}^{0}\},\\
  \mathbf{h}_{u,i}^{0}&=\mathbf{e}_{v_i}+\mathbf{p}_i,
\end{split}
\label{eq.embedding layer}
\end{equation}
where $e_i \in E$ is the representation of item $v_i$, $p_i \in P$ is the representation for  position  index $i$ and $h_{u,i}^0$ is the input representation of $v_i$.
\subsubsection{Self-attention Block} After the embedding layer, we employ the self-attention block to capture the dependencies between items in the sequence and extract the preference of users. Like transformer\cite{vaswani2017attention}. The multi-head self-attention mechanism is adopted to effectively extract selective information in different subspace. Specifically, It first projects the input to $k$ subspace and applies self-attention to each subspace. Then it combines the output of each head by linear projecting the concatenated embedding.  this process is formulated as:
    \begin{gather}
          MH(\mathbf{H}^l)=concat(head_1;head2;,...;head_k)W^O \notag \\
          head_i=Attention(\mathbf{H}^lW_i^Q,\mathbf{H}^lW_i^K,\mathbf{H}^lW_i^V)
    \end{gather}
    where $W_i^Q,W_i^K,W_i^V \in \mathbb{R}^{d\times d \times \frac{d}{h}}$ and $W^O \in \mathbb{R}^{d \times d}$are the learnable project matrix.

In this paper, we adopt the scaled dot-product attention as the attention function:
\begin{gather}
    Attention(Q,K,V)=softmax(\frac{QK^\top}{\sqrt{d/h}})V
\end{gather}
where $K,Q,V$ denote queries, keys and values vector, respectively. And $\sqrt{d/h}$ is the scale factor to avoid large values of inner product.
\subsubsection{Point-Wise Feed-Forward Network}
The multi-head self-attention mechanism is effective in extracting the relation between items, while the simple linear projection limits the capacity of model. To endow the model with nonlinear interactions between different dimensions, we adopt a \emph{Position-wise Feed-Forward Network} after the multi-head self-attention mechanism sub-layer, which is formulated as:
\begin{gather}
    PFFN(\mathbf{H}^l)=[FFN(\mathbf{h}_1^l)^\top,...;FFN(\mathbf{h}_t^l)^\top]^\top \notag \\
    FFN(\mathbf{h}_i^l)=GELU(\mathbf{h}_i^l+b_1)W_2+b2
\end{gather}
where $W_1,b1,W_2,b2$ are learnable parameters. $GELU$ is activation function and is widely adopted in other works. \cite{vaswani2017attention,devlin2018bert,sun2019BERT4Rec}

\subsubsection{Stack more Blocks}. Empirically, stacking more layers is benefit for modeling patterns in deep learning. While model generally becomes difficult to train with it goes deeper. Residual connection, dropout and layer normalization module are proved effectively to alleviate this issue. So we apply them on our model:
    \begin{gather}
        LayerNorm(x+Dropout(sbulayer(x)))
    \end{gather}
where $sublayer$ is the above self-attention block.

In summary, our sequence-based modeling module is represented as follows:
 \begin{gather}
     \mathbf{H}^0=\mathbf{E}+\mathbf{P} \notag \\
     \mathbf{H}^l=Trm(\mathbf{H}^{l-1}) \notag \\
     Trm(\mathbf{H}^{l-1})=LayerNorm(\mathbf{A}^{l-1}+Dropout(PPFN(\mathbf{A}^{k-1}))) \notag \\
     \mathbf{A}^{l-1}=LayerNorm(\mathbf{H}^{l-1},Dropout(MH(\mathbf{H}^{l-1})))
 \end{gather}
 Input a behavior sequence of a user, our multi-layer self-attention blocks output a series of representations of all step $t$. In here, we adopt the last step to output the sequence as the user's local preference under this behavior. We denote it as $\mathbf{e_u}=\bm{h}_{u,|s_{u}|}^{L}$.

\subsection{Discussion on ID-centric Pre-training}
\label{sec:discussion}

Recently, there have been efforts to explore text-based pre-trained recommendation \cite{hou2022towards}, which aims to supplant item ID with natural language. However, we argue that ID information (i.e. ID embedding) is indispensable, which is more suitable to store luxuriant behavior information in recommendation. We attempt to design an ID-centric pre-trained recommendation model for the following reasons:
(1) \emph{\textbf{Natural language is difficult to provide a comprehensive portrayal of an item}}. An item within the recommendation system encompasses a multitude of information, including behavioral logs, visual content, textual description, and so on. Textual information constitutes only a portion of this broader context. Actually, recent studies have corroborated this stance  \cite{gao2023chat,liu2023chatgpt,hou2023large}. These studies demonstrate that even with the help of large language models (LLMs), text-based collaborative filtering struggles to outperform traditional collaborative filtering methods.
(2) \emph{\textbf{ID-based pre-training is far more efficient and low-cost compared to LLM-based models}}. In model serving, our IDP's computation and memory costs are similar to its base sequential encoder's, while LLM-based models require expensive costs. Moreover, in training, CDIM is only functioned for initialization.
(3) \emph{\textbf{ID-based SR models still dominate most real-world systems, and ID-based pre-training is more flexible to be deployed in practice}}. ID-based pre-trained model is more similar to existing SR models, which is easy to deploy and maintain, welcomed by the industry.
Moreover, IDP is universal with different base sequential models, which is more suitable to be enhanced with better sequential modeling (e.g., new architecture or multi-interest modeling). In contrast, existing PLM-based SR models generally bond with specific heavy architectures, which are hard to use future techniques in SR modeling.

\subsection{Complexity analysis}
\label{sec:complexity}
Now, we discuss the complexity of generating ID embeddings for the downstream domain. If we directly search for the k most similar items for all items in the downstream scenarios, the time complexity would be $O(MdN^2log(m))$ (accelerated by heap sort), where $M$ represents the number of downstream scenarios, $N$ denotes the total number of items in the pre-training dataset, $d$ denotes the ID embedding dimension, and $m$ is top-$m$ similar items.  This time complexity is unaffordable, especially when both $M$ and $N$ are considerably large. However, thanks to efficient similarity search algorithms, we can rapidly search out those items. For example, the time complexity can be reduced to $O(M log( log(N))) $ by utilizing the HNSW algorithm \cite{malkov2018efficient}. Thus, our approach is feasible in practice.

\section{Experiments}

In this section, we aim to answer the following questions:
\begin{itemize}
    
 \item  \textbf{ RQ1:}How does IDP perform compared with other state-of-the-art (SOTA) baselines?
 \item \textbf{ RQ2:} Can our proposed framework be generalized to other base models? Furthermore, can our IDP be generalized to other downstream sequential models?
   \item  \textbf{ RQ3:}  How does IDP perform in the zero-shot setting? 
    \item  \textbf{ RQ4:} What are the effects of different modules in IDP? 
    \item \textbf{ RQ5:}  Can our IDP be utilized for helping inner-domain recommendation? 
    \item \textbf{ RQ6:} Can our IDP be extended to multi-modal information?
  
    \item \textbf{ RQ7:}  Can we derive some interesting insights by analyzing IDPs?
\end{itemize}
\begin{table}[!htbp]

\caption{Statistics of all datasets.}
\label{tab:dataset}
\center
\begin{tabular}{l|rrrr}
\toprule
Dataset &\# user&\# item&{\# instance}&{\# avg$_{n}$}\\
\midrule
\multirow{1}{*}{\textbf{Pre\ training}}
~&1,125,233&729,305&13,920,393&19.09\\
\multirow{1}{*}{\ Home}
~&599,247&315,932&6,284,657&19.89\\
\multirow{1}{*}{\ CD}
~&82,891&109,840&1,266,516&11.53\\
\multirow{1}{*}{\ Grocery}
~&102,274&84,303&1,105,163&13.11\\
\multirow{1}{*}{\ Kindle}
~&118,061&140,592&2,188,322&15.57\\
\multirow{1}{*}{\ Movie}
~&222,760&78,638&3,075,735&39.11\\
\midrule
\multirow{1}{*}{\textbf{Downstream}}
\\
\multirow{1}{*}{\ Arts}
~&1,576,189&302,370&2,726,742&9.02\\
\multirow{1}{*}{\ Prime}
~&247,640&10,812&447,359&41.38\\
\multirow{1}{*}{\ Instruments}
~&903,060&112,132&1,469,965&13.11\\
\multirow{1}{*}{\ Office}
~&3,402,597&306,612&5,382,474&17.55\\
\bottomrule
\end{tabular}
\end{table}

\subsection{Dataset}

 We evaluate the proposed method on nine different categories of Amazon review datasets, including pre-training datasets and downstream datasets.
The statistical information of datasets is shown in Table 1.

\textbf{Pre-traning datasets} We select five rating-only  Amazon review datasets as the pre-training datasets including "Grocery and Gourmet Food", "Home and Kitchen", "CDs and Vinyl", "Kindle Store" and "Movies and TV ".
  
\textbf{Downstream datasets}. To evaluate the proposed IDP, we conduct experiments on both cross-domain datasets and cross-platform datasets.
  We select four rating-only  Amazon review datasets as the downstream datasets including "Prime Pantry",  "Musical Instruments", "Arts, Crafts and Sewing" and "Office Products". 

  Following \cite{he2017neural2},  we transformed all rating datasets into the implicit dataset, in which each instance are labeled with 0 or 1 to indicate whether a user rated an item.
whether the user has rated the item
  For the pre-training datasets, following previous works \cite{kang2018self,zhou2020s3}, we discard the users and items with less than five interactions,  as the cold start problem of the pre-training domain is not our concern. We first conduct pre-training on all five pre-trained models. Subsequently, we fine-tune the pre-trained model separately on five downstream datasets to validate the proposed method.
  For all datasets, We adopt the title of products as the description text of items.


\begin{longtable}{l|l|cccccccc|ccr}
\caption{Results on the multi-domain recommendation. The best and the second-best performances are bold and underlined. \textbf{"Impro."} denotes the relative improvement over the best baselines. All improvements are significant (t-test with p ${<}$ 0.05).} \label{tab:main_table} \\

\toprule
Dataset&Model & NDCG@1 & NDCG@3 & NDCG@5 & HR@3 & HR@5 & MRR \\
\midrule
\multirow{11}{*}{Arts}

~ & GRU4Rec & 0.2343 & 0.3346 & 0.3698 & 0.4071 & 0.4928 & 0.3569 \\
~ & SASRec & 0.3228 & 0.4027 & 0.4298 & 0.4602 & 0.5262 & 0.4211 \\
~ & Bert4Rec & 0.2446 & 0.3485 & 0.3837 & 0.4234 & 0.5091 & 0.3691 \\
~ & DACDR & 0.2666 & 0.3601 & 0.3923 & 0.4275 & 0.5059 & 0.3805 \\
~ &PterRec & 0.2713 & 0.3613 & 0.3905 & 0.4259 & 0.4968 & 0.3799 \\
~ & FDSA & 0.2996 & 0.4005 & 0.4341 & 0.4731 & 0.5546 & 0.4219 \\
~ & Recformer & 0.3281 & 0.4137 & 0.4438 & 0.4754 & 0.5486 & 0.4366 \\

~ & UniSrec$_{ID+t}$ & \underline{0.3841} & \underline{0.4926} & \underline{0.5261} & \underline{0.5699} & \underline{0.6514} & \underline{0.5081} \\
\cmidrule{2-8}
~& IDP & 0.3404 & 0.4249 & 0.4528 & 0.4855 & 0.5532 & 0.4423 \\
~& IDP$_{ID+t}$ & \textbf{0.4176} & \textbf{0.5153} & \textbf{0.5454} & \textbf{0.5849} & \textbf{0.6581} & \textbf{0.5302} \\
~ & Impro &8.02\% & 4.41\% & 3.54\% & 2.56\% & 1.02\% & 4.17\%
\\

\midrule
\multirow{11}{*}{Instruments}
~ &GRU4Rec& 0.2382 & 0.3361 & 0.3728 & 0.4073 & 0.4964 & 0.3618 \\
~ &SASRec & 0.3121 & 0.3959 & 0.4243 & 0.4562 & 0.5252 & 0.4149 \\
~ &Bert4Rec & 0.2623 & 0.3648 & 0.4017 & 0.4390 & 0.5286 & 0.3877 \\
~ &DACDR & 0.2774 & 0.3858 & 0.4219 & 0.4640 & 0.5517 & 0.4047 \\
~ &PterRec& 0.2485 & 0.3409 & 0.3747 & 0.4077 & 0.4901 & 0.3655 \\
~ &FDSA & 0.3042 & 0.4108 & 0.4463 & 0.4873 & 0.5737 & 0.4318 \\
~ &Recformer & 0.2812 & 0.3382 & 0.3719 & 0.4015 & 0.4835 & 0.3666 \\
~ &UniSRec$_{ID+t}$ & \underline{0.3404} & \underline{0.4492} & \underline{0.4846} & \underline{0.5274} & \underline{0.6135} & \underline{0.4686} \\
\cmidrule{2-8}
~ &IDP &0.3286 & 0.4207 & 0.4504 & 0.4865 & 0.5588 & 0.4388 \\

~ &IDP$_{ID+t}$& \textbf{0.3977} & \textbf{0.4990} & \textbf{0.5305} & \textbf{0.5715} & \textbf{0.6481} & \textbf{0.5149} \\
~ &Impro. & 14.41\% & 9.98\% & 8.65\% & 7.72\% & 5.34\% & 8.99\%\\

\midrule
\multirow{11}{*}{Office}
~ & GRU4Rec & 0.2884 & 0.4052 & 0.4460 & 0.4896 & 0.5887 & 0.4267 \\
~ &SASRec & 0.3408 & 0.4429 & 0.4790 & 0.5169 & 0.6047 & 0.4619 \\
~ &Bert4Rec & 0.3101 & 0.4296 & 0.4681 & 0.5155 & 0.6089 & 0.4467 \\
~ &DACDR & 0.2879 & 0.4054 & 0.4462 & 0.4907 & 0.5896 & 0.4245 \\
~ &PterRec & 0.3082 & 0.4155 & 0.4501 & 0.4927 & 0.5768 & 0.4329 \\
~ &FDSA & 0.3317 & 0.4458 & 0.4832 & 0.5278 & 0.6185 & 0.4645 \\
~ &Recformer & 0.2521 & 0.3142 & 0.3382 & 0.3590 & 0.4176 & 0.3429 \\
~ &UniSRec$_{ID+t}$ & \underline{0.3899} & \underline{0.5055} & \underline{0.5418} & \underline{0.5882} & \underline{0.6763} & \underline{0.5205} \\
\cmidrule{2-8}
~ &IDP &0.3458 & 0.4487 & 0.4831 & 0.5227 & 0.6088 & 0.4668 \\
~ &IDP$_{ID+t}$ & \textbf{0.4096} & \textbf{0.5186} & \textbf{0.5528} & \textbf{0.5965} & \textbf{0.6798} & \textbf{0.5331} \\
~&Impro & 4.81\% & 2.53\% & 1.99\% & 1.39\% & 0.51\% & 2.36\%\\
\midrule
\multirow{11}{*}{Prime}
~ & GRU4Rec & 0.1094 & 0.1822 & 0.2163 & 0.2361 & 0.3193 & 0.2219 \\
~ & SASRec & 0.1520 & 0.2224 & 0.2536 & 0.2743 & 0.3503 & 0.2584 \\
~ & Bert4Rec & 0.0950 & 0.1561 & 0.1861 & 0.2013 & 0.2747 & 0.1977 \\
~ & DACDR & 0.1192 & 0.1896 & 0.2214 & 0.2415 & 0.3191 & 0.2268 \\
~ & PterRec & 0.1119 & 0.1804 & 0.2136 & 0.2314 & 0.3123 & 0.2209 \\
~& FDSA  & 0.1209 & 0.1877 & 0.2188 & 0.2370 & 0.3127 & 0.2261 \\
~ & Recformer & 0.1519 & 0.2069 & 0.2298 & 0.2450 & 0.3010 & 0.2395 \\
~ &UniSRec$_{ID+t}$ & 0.1280 & 0.1991 & 0.2313 & 0.2518 & 0.3302 & 0.2372 \\
\cmidrule{2-8}
~ & IDP & \underline{0.1526} & \underline{0.2268} & \underline{0.2595} & \underline{0.2804} & \underline{0.3598} & \underline{0.2629} \\
~ & IDP$_{ID+t}$ & \textbf{0.1842} & \textbf{0.2630} & \textbf{0.2953} & \textbf{0.3209} & \textbf{0.3996} & \textbf{0.2965} \\
~ & Impro. &21.18\% & 16.99\% & 16.44\% & 16.99\% & 14.07\% & 16.37\%\\
\bottomrule
\end{longtable}

\subsection{Experimental Setting}
\textbf{Implement Details.} The embedding sizes of items are set to 64 and the batch size is 256 for all methods. We optimize all models with the Adam Optimizer and carefully search for the hyper-parameters of all baselines. To avoid overfitting,  we adopt the early stop strategy with a patience of 20 epochs. We adopt the random negative sampling strategy \cite{rendle2009bpr} with 1 sampling number.  For fair comparison, we employed the BPR loss for all baseline models, as we observed that BPR loss outperformed cross-entropy loss in performance.
We conduct a grid search for hyper-parameters. We search for models’ learning rates among $\{1e^{-3}, 3e^{-4}, 1e^{-4}, 3e^{-5}\}$.

\noindent
\textbf{Evaluation Protocols.} 
We adopt the leave-one-out strategy to evaluate the models' performance. Following~\cite{kang2018self,zhou2020s3},  for each ground truth, we randomly sample $99$ items that the user did not interact with under the target behavior as negative samples.  
We employ top-K hit rate (HR@K), top-K normalized discounted cumulative gain (NDCG@K), and mean Reciprocal Rank ( MRR), with $k=\{1,3,5\}$.

 Specifically, the HR@k is calculated by the following: 
 \begin{equation}
    \begin{split}
    HR@k=\frac{1}{|U|}\sum_{u\in U} \delta(\hat{R}(u,k) \cap R(u) \neq \emptyset),
    \end{split}
\label{HIT}
\end{equation}
where $\hat{R}(u,k)$ is the top-k predicted item list of user $u$ that sorting by predicted score. $R(u)$ is a ground-truth set of items of user $u$. $\delta(\cdot)$ is the indicator function, where the $\delta(b)=1$ if $b$ is true and 0 if false.

\noindent The $NDCG@k$  calculated by the following:  
 \begin{equation}
  \begin{aligned}
&\qquad\operatorname{rel}(u,i) =\delta({\hat{R}(u,k)}^i \in R(u)), \\
\operatorname{NDCG}(N) &=\frac{1}{Y} \sum_{i=1}^{N} \frac{2^{\operatorname{rel}(u,i)}-1}{\log (i+1)} \text { where } Y=\sum_{i=1}^{N} \frac{1}{\log (i+1)}, \\
\end{aligned}
\label{eq.NDCG}
\end{equation}
where ${\hat{R}(u,k)}^i$ is the $i$-th item of ranked item list $\hat{R}(u,k)$. $\delta(\cdot)$ is the indicator function.

\noindent The $MRR$ is calculated by the following: 
 \begin{equation}
    \begin{split}
    MRR=\frac{1}{|U|}\sum_{u \in U} rank_{u}
    \end{split}
\label{eq.MRR}
\end{equation}

$rank_{u}$ is the rank position of the first relevant item  in $\hat{R(u)}$.

 \subsection{Baselines}

\label{Baselines}
  To validate our IDP, we conduct a comprehensive comparison. Specifically, we compare our method with the following baselines:(1)\textbf{ GRU4Rec}\cite{hidasi2016session} (2)\textbf{ SASRec} \cite{kang2018self}, (3)\textbf{ BERT4Rec} \cite{sun2019BERT4Rec}, (4)\textbf{ FDSA} \cite{zhang2019feature}, (5)\textbf{Recformer}\cite{li2023text}, and (6)\textbf{UniSRec}$_{ID+t}$ \cite{hou2022towards}. 

 \begin{itemize}[leftmargin=*]
       \item { GRU4Rec} \cite{hidasi2016session}, which adopts GRU as users' sequence encoder for sequential recommenadtion. 
     \item{ SASRec} \cite{kang2018self}, which is a classical sequential recommendation model that introduces self-attention in behavior modeling. 
     \item{ BERT4Rec} \cite{sun2019BERT4Rec}, which adopts masked item prediction as an optimizing object.
     \item{ FDSA} \cite{zhang2019feature}, which adopts self-attentive networks to model item and feature transition. In this paper, we adopt item descriptions as item features. 
     \item {DACDR} \cite{zhang2021deep},which is a cross-domain recommendation model. It achieves knowledge transfer across domains by embedding alignment through common features. In this paper, we adopt text information as common feature.
     \item {PeterRec} \cite{yuan2020parameter}, which is a parameter-efficient transfer learning recommendaiton model. It inserts model patches into pre-trained models to achieve effective transfer learning for recommendation.
     \item {Recformer}\cite{li2023text}, which is a pure language model-based sequential recommendation model. Recformer first translates user behavior sequence to a natural language sentence. Then Recformer encodes translated sentences by language model.   
     \item{UniSRec}$_{ID+t}$ \cite{hou2022towards}, which is a SOTA text-based pre-train model. In the pre-training stage, UniSRec represents items with text embedding encoded by a pre-trained language model. Then it encodes user sequential behaviors by a sequential model based on text embedding. In the downstream domain, UniSRec fine-tunes the pre-trained model by the designed MoE-enhanced adopter. UniSRec$_{ID+t}$ is the final version of UniSRec, which adds ID embedding in downstream domains.  
 
 \end{itemize}


 For our model, we adopt SASRec as our base model and pre-train SASRec on pre-training datasets. Considering that some baselines adopt textual vectors as an auxiliary feature (i.e. UniSRec$_{ID+t}$ and FDSA), we also adopt textual vectors as an additional feature for a fair comparison. Following UniSRec, we also add ID embedding with the corresponding textual vector and input it to the sequential encoder. The ID-only version is denoted as IDP and the text-enhanced version is denoted as IDP$_{ID+t}$. 
\subsection{Overall Performance Comparison (RQ1)}
The main results of our IDP are shown in Table \ref{tab:main_table}. We have:
 
(1) Compared to pure ID-based models (i.e., SASRec, GRU4Rec, BERT4Rec), our IDP (adopting SASRec as the base model) significantly outperforms them on all four datasets, demonstrating that our IDP effectively utilizes the capabilities of ID-based pre-trained models. With the assistance of the proposed CDIM, our IDP can use textual information associated with items as a bridge to generate ID embeddings for downstream domains. As a result, our IDP can flexibly transfer knowledge from the pre-trained model's ID embeddings to new domains. Furthermore, comparing text-based models (i.e. Recformer) with ID-based models, the text-based models outperformed the ID-based models on the Arts dataset. However, on other datasets, they performed less effectively than the ID-based models. This indicates textual information still can not replace the ID-based model. 

(2) We note that text-enhanced models achieve better performance than ID-only models. This can be attributed to that the textual information of items serves as a useful auxiliary feature on those datasets.  Furthermore, compared with the text-enhanced method (i.e. FDSA and UniSRec), our IDP$_{ID+t}$ still significantly outperforms the SOTA pre-trained model UniSRec on all datasets. Those results show the superiority of our IDP.   UniSRec is a text-based pre-trained model that transfers textual modality information to downstream domains. In contrast to UniSRec, our IDP is an ID-based pre-trained model.  These experimental results further validate our opinion that modality information can not replace ID embedding to store behavioral information in recommendation. 

(3) In comparison with other baselines, our model achieves substantial improvements when confronted with downstream datasets of reduced scale. Actually, in scenarios where the target domain data is relatively scarce (e.g., Prime), knowledge from other domains is more needed. This demonstrates our IDP can more effectively leverage the knowledge contained in the pre-trained model, especially when the downstream domain lacks sufficient behavioral information for items.



\subsection{Universality of IDP (RQ2)}
We validate the universality of our IDP from two perspectives: (1)the universality of IDP on the pre-training base model. (2)the universality of IDP on the downstream sequential model. (3) the universality of IDP on the cross-platform scenarios.
\subsubsection{The universality of IDP on the pre-training base model.}
As discussed in the Introduction Section, our IDP transfers ID embedding to downstream domains and thus does not depend on a specific sequential recommendation model. To evaluate this, we change our pre-training base model to other ID-based sequential models (i.e., BERT4Rec) and evaluate the performance of pre-trained models on downstream datasets. The results are shown in Table \ref{tab:Universality}.

From this table, we find that (1) our IDP still works well on all datasets when we change the pre-training base model. It shows the universality of our IDP, which does not depend on a specific base model.  In fact, notwithstanding that the base models have different architectures, all of them are able to learn the knowledge within the pre-training datasets. Our IDP serves as a mechanism for facilitating knowledge transfer from the upstream to the downstream dataset. 
(2) Similar to adopting SASRec, our IDP gets more significant improvement when the dataset is smaller.

\begin{table}[!htbp]

\caption{Results of IDP with other sequential model in pre-training stage. IDP$^ {BERT4Rec}$ adopts BERT4Rec as the base SR model both in pre-training stage and fine-tuning stage. All improvements are significant (p ${<}$ 0.05).}
\label{tab:Universality}
\center
\begin{tabular}{l|l|cc}
\toprule
Dataset & Model &HR@5& NDCG@5  \\
\midrule
\multirow{2}{*}{Instruments}
~ &BERT4Rec &0.5286&0.4017\\
~ &IDP$^{BERT4Rec}$ &0.5563&0.4438\\
\midrule
\multirow{2}{*}{Arts}
~ &BERT4Rec &0.5091&0.3837\\
~ &IDP$^{BERT4Rec}$ &0.5098&0.3968\\
\midrule
\multirow{2}{*}{Office}
~ &BERT4Rec &0.6089&0.4681\\
~ &IDP$^{BERT4Rec}$ &0.6029&0.4687\\
\midrule
\multirow{2}{*}{Prime}
~ &BERT4Rec &0.2747&0.1861\\
~ &IDP$^{BERT4Rec}$ &0.3466&0.2422\\
\bottomrule
\end{tabular}
\end{table}

\subsubsection{The universality of IDP on the downstream sequential model.}
As introduced before, in real recommender systems, diverse teams tend to adopt different sequential models based on their business requirements and system architecture.  The coupling of upstream and downstream sequential models (what the previous works adopt) constrains the flexibility of utilizing pre-trained recommendation models. This is one of the motivations why we try to propose an ID-centric Pre-trained Model.
  
In this experiment, we adopt different sequential models for pre-training and downstream fine-tuning.  Specifically, we initially employ SASRec as the base model to pre-train the ID embeddings. Subsequently, in the downstream phase, we utilize BERT4Rec as the sequential model and fine-tune the generated ID embedding. Here, we train BERT4Rec from scratch.
The results are shown in Table \ref{tab:Flexibility}. From the table, we can find that our IDP still can improve downstream performance even when the sequential model is different from upstream. It shows the flexibility of our IDP, which does not bond with the pre-trained sequential encoder. This characteristic makes IDP more likely to be widely 
applied in the real world.

\begin{table*}[!hbpt]

\caption{A challenging setting that uses different SR models in pre-training/tuning. IDP$^{B\&S}$ adopts SASRec as the base SR model in pre-training and uses a different BERT4Rec for fine-tuning. All improvements are significant (p ${<}$ 0.05).}
\label{tab:Flexibility}
\center
\begin{tabular}{l|l|cc}
\toprule
Dataset & Model &HR@5& NDCG@5  \\
\midrule
\multirow{2}{*}{Instruments}
~ &BERT4Rec &0.5286&0.4017\\
~ &IDP$^{B\&S}$ &0.5428&0.4299\\
\midrule
\multirow{2}{*}{Arts}
~ &BERT4Rec &0.5091&0.3837\\
~ &IDP$^{B\&S}$ &0.5149&0.3999\\
\midrule
\multirow{2}{*}{Office}
~ &BERT4Rec &0.6089&0.4681\\
~ &IDP$^{B\&S}$ &0.6130&0.4773\\
\midrule
\multirow{2}{*}{Prime}
~ &BERT4Rec &0.2747&0.1861\\
~ &IDP$^{B\&S}$ &0.3444&0.2406\\
\midrule
\end{tabular}
\end{table*}
  \subsubsection{The universality of IDP on the cross-platform scenarios}

 \begin{table*}[!htbp]
   
\caption{Results on the cross-platform scenario. The best and the second-best performances are bold and underlined. \textbf{"Impro."} denotes the relative improvement over the best baselines. All improvements are significant (t-test with p ${<}$ 0.05).}
\label{tab:cross-platfrom}

\center
\begin{tabular}{l|cccccccc|ccr}
\toprule
Metric & NDCG@1 & NDCG@3 & NDCG@5 & HR@3 & HR@5 & MRR \\
\toprule
GRU4Rec & 0.3374 & 0.4794 & 0.5253 & 0.5808 & 0.6919 & 0.4945 \\
SASRec & 0.3566 & 0.4948 & 0.5362 & 0.5927 & 0.6930 & 0.5082 \\
BERT4Rec & 0.3258 & 0.4668 & 0.5117 & 0.5682 & 0.6765 & 0.4821 \\
DACDR & 0.1444 & 0.2278 & 0.2649 & 0.2891 & 0.3791 & 0.2675 \\
PterRec & 0.2187 & 0.3475 & 0.3961 & 0.4417 & 0.5599 & 0.3763 \\
FDSA & 0.3313 & 0.4667 & 0.5107 & 0.5644 & 0.6710 & 0.4840 \\
Recformer & 0.1332 & 0.2256 & 0.2685 & 0.2942 & 0.3987 & 0.2629 \\
UniSRec$_{ID+t}$ & 0.2902 & 0.4328 & 0.4810 & 0.5361 & 0.6531 & 0.4513 \\
IDP & \textbf{0.3606} & \textbf{0.4966} & \textbf{0.5381} & \textbf{0.5935} & \underline{0.6942} & \textbf{0.5105} \\
IDP$_{ID+t}$ & \underline{0.3566} & 0.4904 & \underline{0.5363} & 0.5874 & \textbf{0.6987} & 0.5062 \\
Impro. & 1.10\% & 0.36\% & 0.35\% & 0.10\% & 0.80\% & 0.45\% \\
\bottomrule
\end{tabular}

\end{table*}
  To demonstrate the universality of our IDP, we conduct a more challenging cross-platform experiment.  We pre-traine the IDP on the Amazon dataset and fine-tune it on the Movielens-1M dataset. The results are shown in Table.\ref{tab:cross-platfrom}.  From the table, we can observe our IDP still achieved an improvement even in cross-platform scenarios. Note that, we argue that it is a challenging task as Amazon and Movielens are different types of platforms (shopping and video websites), thus users may have different behavior patterns in them. It shows the possibility of building a universal ID-centric pre-trained model.
  
\subsection{Results of Zero-shot Scenario (RQ3)}
 
In this section, we further evaluate the effectiveness of our IDP in a zero-shot setting. In this setting, we directly apply the pre-trained model to downstream datasets without tuning. Note that, as the UniSRec$_{ID+t}$ can not be adopted for zero-shot setting.  We compare our method with the zero-shot version of UniSRec (i.e. UniSRec$_t$). We demonstrate the results in Table.\ref{tab:zero-shot}. Note that, for IDP$_{ID+t}$, we employ PCA to reduce the dimension with the encoded textual vectors to the same dimension with ID embeddings. Following UniSRec, we simply add textual vectors with generated ID embedding as the input of the sequential model.

From this table, we can see that Our IDP is indeed effective. Even in zero-shot scenarios, it still can help the system make decisions. Because our IDP can transfer knowledge in the pre-training domains to other domains.  Furthermore, with the help of textual information, our IDP exhibits enhanced performance and outperforms UniSRec by a substantial margin. This is attributed to the universal nature of semantic information across different domains and it is an important auxiliary information. Note that, different from UniSRec which pre-trains an adapter to encode textual vectors, textual vectors are directly used in IDP (i.e. reducing dimension with PCA). This also highlights the flexibility of our IDP.
 
\begin{table}[!htbp]

\caption{Results on the zero-shot setting. All improvements are significant over baselines (t-test with p ${<}$ 0.05).}
\label{tab:zero-shot}
\center
\begin{tabular}{l|l|cc}
\toprule
Dataset & Model &HR@5& NDCG@5  \\
\midrule
\multirow{2}{*}{Arts}
~ &UniSRec$_t$ &0.2620&0.1920\\
~ &IDP &0.1079&0.0681\\
~ &\textbf{IDP $_{ID+t}$} &\textbf{0.3234}&\textbf{0.2608}\\
\midrule
\multirow{2}{*}{Instruments}
~ &UniSRec$_t$ &0.1693&0.1117\\
~ &IDP &0.0799&0.0489\\
~ &\textbf{IDP $_{ID+t}$} &\textbf{0.2648}&\textbf{0.2002}\\
\midrule
\multirow{2}{*}{Office}
~ &UniSRec$_t$ &0.2189&0.1556\\
~ &IDP &0.1337&0.0849\\
~ &\textbf{IDP$_{ID+t}$} &\textbf{0.2708}&\textbf{0.2177}\\
\midrule
\multirow{2}{*}{Prime}
~ &UniSRec$_t$ &0.1598&0.1092\\
~ &IDP &0.0839&0.0519\\
~ &\textbf{IDP $_{ID+t}$} &\textbf{0.1952}&\textbf{0.1487}\\
\bottomrule
\end{tabular}
\end{table}

 \subsection{Ablation Study (RQ4)}
 
In this section, we conduct an ablation study to evaluate the effectiveness of the proposed modules of our IDP. Precisely, we have (a) IDP w/o M, which removes our matcher CDIM and only fine-tunes the pre-trained ID-based sequential encoder on downstream domains, (b) IDP w/o BCL, which directly utilizes the original SimCSE model as the cross-domain ID matcher without the behavior-based CL tasks.  

\begin{figure}[!hbtp]
\centering
\includegraphics[width=0.79\textwidth]{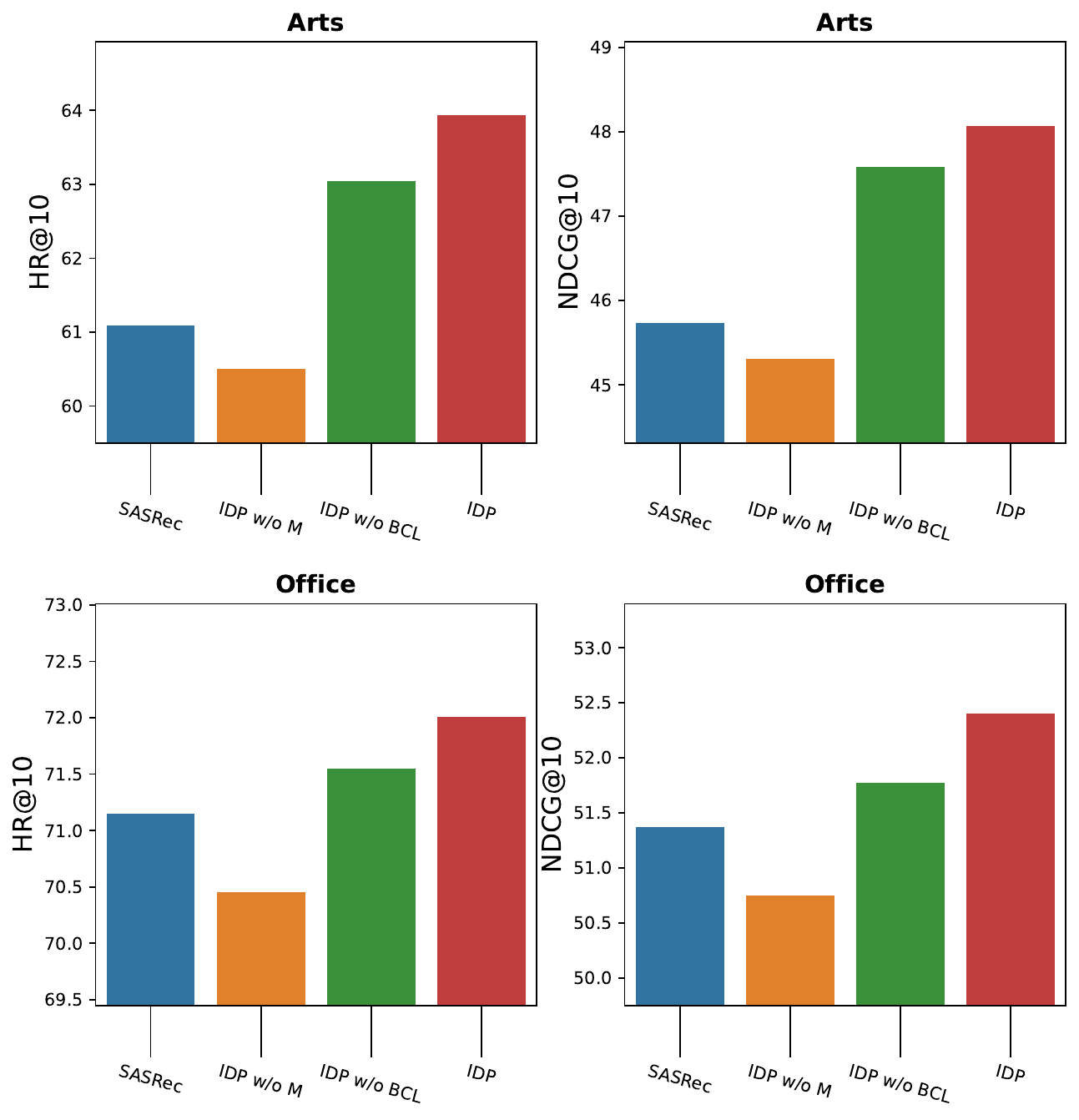}
\caption{Ablation studies on Arts and Office datasets. All components of IDP are effective.}
\label{fig:ablation}
\end{figure}

The results are reported in Fig.~\ref{fig:ablation}. We find 
(1) By comparing  IDP with IDP w/o M, we can see that our proposed CDIM can effectively transfer knowledge contained in the pre-trained model via ID embeddings, which utilizes modality information as a bridge. Besides, IDP w/o M even performs worse than the single-domain SASRec. This indicates that only transferring the pre-trained sequential encoder may lead to negative transfer effects.  
(2) By comparing IDP with IDP without BCL, we can observe that the designed behavior-aware contrastive learning task indeed functions effectively. It accomplishes this by integrating behavioral information into textual vectors through the tuning of SimCSE. As a result, CDIM is able to balance behavioral and textual similarity across domains, facilitating the aggregation of ID embeddings for initialization.

\subsection{Representing Intra-domain Cold-start Items via Pre-trained Item Embeddings (RQ5)}

\begin{table}[!htbp]

\caption{The result of generating ID embedding within inner-domain.}

\label{tab:inner_domain}

\center

\begin{tabular}{l|l|cccccc}
\toprule
Dataset & Metric&HR@3 &NDCG@3 & HR@5 & NDCG@5   \\
\midrule
\multirow{1}{*}{Instruments}
~ &IDP$^{Inner}$&0.1950&0.1655 &0.2352 &0.1819 \\
\midrule
\multirow{1}{*}{Arts}
~ &IDP$^{Inner}$&0.1294 &0.1053 &0.2705&0.1874\\
\bottomrule
\end{tabular}

\end{table}
In this section, we aim to investigate the effects of generating IDs within the inner domain. This experimental result has the ability to demonstrate whether our IDP is capable of addressing inner-domain cold start issues. Instead of generating ID embeddings from the upstream domain to the downstream domain, we focus on generating ID embeddings within the inner domain. 
Specifically, for a domain $A$, we first train a sequential model. Then we use our IDP to generalize ID embedding for cold items (without any interaction).  Last we directly evaluate the model's performance on cold items without any tuning operation. Here, we randomly select 10\% items as cold items and remove them from the training dataset. The results are shown in Table \ref{tab:inner domain}, where IDP$ ^{Inner}$ is the model that re-generates ID embedding by our CDIM  in the Inner domain. From this table, we can observe that our IDP can still provide reasonably accurate recommendations.  This indicates that our IDP can be flexibly applied to inner-domain item cold start scenarios. 

\subsection{Extending to Multi-modal Scenario}
\begin{table*}[!htbp]
\caption{Results on of extending textual modality information to multi-modal information. The best performances are bold and underlined.  All improvements are significant (t-test with p ${<}$ 0.05).}
\label{tab:extending MM}

\center

\begin{tabular}{l|l|cccccccc|ccr}

\toprule
Dataset&Model & NDCG@1 & HR@3 & NDCG@3 & HR@5 & NDCG@5 & MRR \\
\midrule
\multirow{6}{*}{Arts}
~ &SASRec &0.3228&0.4602&0.4027&0.5262&0.4298&0.4211\\
~ &UniSRec &0.3841&0.5699&0.4926&0.6514&0.5261&0.5081\\
~ &IDP &0.3404&0.4855&0.4249&0.5532&0.4528&0.4423\\
~ &IDP$_{ID+t}$ &0.4176&0.5849&0.5153&0.6581&0.5454&0.5302\\
~ &IDP$^{MM}$ &0.3453&0.4860&0.4272&0.5493&0.4533&0.4436\\
~ &IDP$^{MM}_{ID+t}$ &\textbf{0.4185}&\textbf{0.5871}&\textbf{0.5169}&\textbf{0.6607}&\textbf{0.5472}&\textbf{0.5315}\\
\midrule
\multirow{6}{*}{Office}
~ &SASRec &0.3408&0.5169&0.4429&0.6047&0.4790&0.4619\\
~ &UniSRec &0.3899&0.5882&0.5055&0.6763&0.5418&0.5205\\
~ &IDPID &0.3458&0.5227&0.4487&0.6088&0.4831&0.4668\\
~ &IDP$_{ID+t}$ &\textbf{0.4096}&\textbf{0.5965}&\textbf{0.5186}&\textbf{0.6798}&\textbf{0.5528}&\textbf{0.5331}\\
~ &IDP$^{MM}$ &0.3461&0.5242&0.4497&0.6087&0.4845&0.4681\\
~ &IDP$^{MM}_{ID+t}$ &0.4056&0.5944&0.5156&0.6783&0.5502&0.5302\\
\midrule
\multirow{6}{*}{Instruments}
~ &SASRec &0.3121&0.4562&0.3959&0.5252&0.4243&0.4149\\
~ &UniSRec &0.3404&0.5274&0.4492&0.6135&0.4846&0.4686\\
~ &IDPID &0.3286&0.4865&0.4207&0.5588&0.4504&0.4388\\
~ &IDP$_{ID+t}$ &0.3977&0.5715&0.4990&0.6481&0.5305&0.5149\\
~ &IDP$^{MM}$ &0.3326&0.4867&0.4224&0.5568&0.4512&0.4401\\
~ &IDP$^{MM}_{ID+t}$ &\textbf{0.3998}&\textbf{0.5024}&\textbf{0.5341}&\textbf{0.5756}&\textbf{0.6525}&\textbf{0.5181}\\
\midrule
\multirow{6}{*}{Prime}
~ &SASRec &0.1520&0.2743&0.2224&0.3503&0.2536&0.2584\\
~ &UniSRec &0.1280&0.2518&0.1991&0.3302&0.2313&0.2372\\
~ &IDPID &0.1539&0.2804&0.2268&0.3598&0.2595&0.2629\\
~ &IDP$_{ID+t}$ &0.1842&0.3209&0.2630&0.3996&0.2953&0.2965\\
~ &IDP$^{MM}$ &0.1522&0.2825&0.2273&0.3616&0.2598&0.2628\\
~ &IDP$^{MM}_{ID+t}$ &\textbf{0.1873}&\textbf{0.3213}&\textbf{0.2645}&\textbf{0.4012}&\textbf{0.2973}&\textbf{0.2986}\\

\bottomrule
\end{tabular}

\end{table*}
 In this paper, we aim to utilize the modality information to train our CDIM and transfer the ID embedding to downstream domains. In the Sec. \ref{secCDIM} we mainly utilize the textual modality information as a bridge as it is more commonly available in the real world.  However, our IDP can easily be extended to multi-modal scenarios.  To evaluate it, we both utilize the image information and textual information to train the CDIM. Specifically, we first utilize pre-trained VIT\cite{dosovitskiy2020image} to extract the image feature $im_v$ of items. Then we fuse the encoded textual feature $t_v$  and image feature $m_v$ by a two-layer MLP with a dropout layer, which is represented as $mm_v$. Subsequently, we replace the textual representation $t_v$ in Eq.\ref{eq.SimCSE_textual} and Eq.\ref{eq.SimCSE_behavior} with multi-modal representation $mm_v$, and conduct the modality information contrastive learning (i.e., Eq.\ref{eq.SimCSE_textual}) and behavior involved tuning (i.e., Eq.\ref{eq.SimCSE_behavior}). After that, we get a multi-modal-based CDIM, which can be used to transfer pre-trained ID embedding downstream.

 To evaluate the performance of our IDP in a multi-modal scenario, we conduct the following experiments (1) IDP$^{mm}$, which utilize the multi-modal based CDIM  as the matcher (2) IDP$^{mm}_{ID+mm}$, which is an multi-modal enhanced IDP and utilize the multi-modal representation as an auxiliary feature.  

 We show the results in the Table.\ref{tab:extending MM}. From this table, we have the following observations: \textbf{(1)} All variants of our IDP outperform SASRec and all various of our textual-enhanced IDP outperform UniSRec, which further demonstrates the power of our IDP. \textbf{(2)} By comparing IDP and IDP$^{MM}$, we found that introducing multimodal information to train the CDIM module indeed helps improve the performance of our IDP. (3) By further comparing C and D, we found that in the case of text augmentation, C generally outperforms D. This further indicates the benefit of introducing multimodal information to our IDP. Furthermore, This experiment demonstrates the versatility of our IDP, which can be applied in multi-modal scenarios.

\subsection{Analysis of Sequential Encoder (RQ6)}
In this paper, we propose the concept of ID-centric pre-trained models for the first time. In this section, we conduct two exploratory experiments aimed at providing some insights.

\textbf{The effect of the pre-trained sequential encoder.}
Here, we aim to further validate the impact of the pre-trained sequential encoder. Specifically, instead of both utilizing a pre-trained encoder and generated ID-embedding for downstream fine-tuning, we only adopt generated ID-embedding for downstream fine-tuning and train the sequential encoder from scratch, denoted as \textbf{IDP w/o PS}. We report our results in Table \ref{tab:withoutencoder}. We can observe that IDP performs worse than IDP w/o PS in the Office dataset, whereas the performance is reversed in the Prime dataset. This suggests that the utilization of a pre-trained sequential encoder could potentially lead to negative transfer in specific scenarios. In such cases, training the sequential encoder from scratch might be a more suitable option.
  

\begin{table}[!htbp]

\caption{Performance comparison of IDP and IDP w/o PS.}
\label{tab:withoutencoder}

\center

\begin{tabular}{l|l|cccccc}
\toprule
Dataset & Metric & HR@3&NDCG@3&HR@5 & NDCG@5   \\
\midrule
\multirow{2}{*}{Instruments}
~ &IDP&\textbf{0.4865}&\textbf{0.4207} &\textbf{0.5588} &\textbf{0.4504} \\
~ &IDP w/o PS &0.4843&0.4220&0.5522&0.4500\\
\midrule
\multirow{2}{*}{Prime}
~ &IDP &0.2804&0.2268&0.3598&0.2595\\
~ &IDP w/o PS&\textbf{2912}&\textbf{0.2364}&\textbf{0.3715}&\textbf{0.2693}\\
\bottomrule
\end{tabular}

\end{table}
\begin{table*}[!htbp]
\caption{The result of exchanging sequential encoders. SASRec$^{EX}$ is the version of SASRec with the exchanged encoder.}
\label{tab:changeEncoder}

\center

\begin{tabular}{l|l|cccccc}
\toprule
Dataset & Metric &HR@3 &NDCG@3 & HR@5 & NDCG@5  \\
\midrule
\multirow{2}{*}{Instruments}
~&SASRec &0.4562&0.3959&0.5252&0.4243\\
~& SASRec$^{EX}$&0.4191&0.3553&0.4903&0.3846\\
\midrule
\multirow{2}{*}{Prime}
~&SASRec &0.2743&0.2224&0.3503&0.2536\\
~& SASRec$^{EX}$&0.2259&0.1810&0.2946&0.2092\\
\bottomrule
\end{tabular}

\end{table*}

\textbf{The transferability of the sequence encoder.}
To analyze the transferability of the sequential model's encoder, we perform an exchange of the sequential encoders between two sequential models that are trained on different domains. Specifically, we train two sequential models on the "Musical Instruments" (Instruments) and "Prime Pantry" (Prime) datasets, respectively. Afterward, we interchange the sequential encoders of these two models while retaining their respective embedding layers. 
We directly evaluate the performance of those models on the test dataset without any fine-tuning operation.
The results are reported in Table \ref{tab:changeEncoder}, where SASRec$^{EX}$ represents the model with the exchanged encoder.

From this table, it is evident that SASRec$^{EX}$ does not perform as well as SASRec, which is an expected outcome since the encoder of SASRec$^{EX}$ is trained on a different domain. Nevertheless, SASRec$^{EX}$ still demonstrates respectable performance on both the Instrument and Prime datasets, despite the performance gap compared to SASRec. It's worth noting that there is a weak correlation between "prime" and "Instrument."

Based on these findings, we hypothesize that sequence models might contain limited domain-specific knowledge, with the primary knowledge being stored within ID embeddings. The encoder may primarily serve to capture the patterns represented by user behavior sequences in a domain-agnostic manner. This indicates the importance of knowledge transfer from ID embeddings when designing an advanced pre-trained model.

\section{Conclusion And Future Work}

In this work, we propose an ID-centric recommendation pre-training framework (IDP) for MDR, which can effectively transfer the knowledge to downstream domains. IDP directly aggregates well-learned ID embeddings in pre-training to build new items' initializations via the behavior-aware CDIM, which is effective, universal, and flexible to be deployed with most existing recommendation models in practice. In the future, we will consider utilizing multi-modal information for knowledge cross-domain transfer.
\bibliographystyle{ACM-Reference-Format}
\bibliography{Main}
\end{document}